# Temperature effect on non-Darcian flow in low-permeability porous media

Yuntian Teng[a], Yifeng Wang[b], Zihao Li[c], Rui Qiao[d], Cheng Chen[a,1]

[a] Department of Civil, Environmental, and Ocean Engineering, Stevens Institute of Technology, Hoboken, NJ 07310

[b] Sandia National Laboratories, Albuquerque, NM 87185

[c] School of Civil and Environmental Engineering, Cornell University, Ithaca, NY 14853

[d] Department of Mechanical Engineering, Virginia Tech, Blacksburg, VA 24060

## Abstract

In low-permeability porous media, the velocity of a fluid flow exhibits a nonlinear dependence on the imposed pressure gradient. This non-Darcian flow behavior has important implications to geological disposal of nuclear waste, hydrocarbon extraction from shale, and flow and transport in clay-rich aquifers. Temperature has been postulated to affect the threshold pressure gradient of a non-Darcian flow; however, the supporting data is very limited. In this study we for the first time report a systematic measurement of the threshold pressure gradient under various permeabilities and temperatures. The results show that a higher temperature leads to a lower threshold pressure gradient under the same permeability and a faster reduction of the threshold pressure gradient with increasing permeability. The experimental data are fitted to a two-parameter model to determine the parameters, $h_0$ and $a$, which characterize the interfacial fluid-solid interactions and the transition between the Darcy and non-Darcian regimes.

[1] To whom correspondence may be addressed. Email: cchen6@stevens.edu

## Plain Language Summary

Darcy's law specifies a linear relation between the velocity of a fluid flow and the imposed pressure gradient in a porous medium. However, in low-permeability porous media a threshold pressure gradient is required to trigger a fluid flow. This phenomenon is referred to as non-Darcy flow, which has important implications to many natural and engineered processes in low-permeability porous media, such as geological disposal of nuclear waste, hydrocarbon extraction from shale, and flow and transport in clay-rich aquifers. Temperature has been postulated to affect the threshold pressure gradient of a non-Darcian flow; however, the supporting data is very limited.

In this study, for the first time, we systematically investigate the temperature effect on the transition between Darcy and non-Darcian flow regimes. The results show that a higher temperature leads to a lower threshold pressure gradient and a faster reduction of the threshold pressure gradient with increasing permeability. The experimental data are then fitted to a two-parameter model to characterize the interfacial fluid-solid interactions and the transition between the Darcy and non-Darcian regimes. The experimental data and the model allow for a systematic prediction of fluid flow for a general set of low-permeability porous media under various temperatures and pressure gradients.

## 1. Introduction

The widely used Darcy's law specifies a linear relation between the Darcy velocity of a fluid flow and the pressure gradient that drives the flow (1). However, many studies (1-8) have shown that Darcy velocity can exhibit a nonlinear dependence on the pressure gradient in low-permeability porous media such as clay and shale when the pressure gradient is adequately low. This phenomenon is referred to as non-Darcian flow and is caused by strong liquid-solid interactions near solid surfaces (2), due to the combined effects of various interfacial and intermolecular forces including the van der Waals forces (9). In low-permeability porous media, the nanoscale pore size is comparable to the interfacial layer thickness so that the influence of strong liquid-solid interactions is not negligible (10). In this circumstance, a threshold pressure gradient is required to overcome the resistance from the interfacial layer to trigger a fluid flow (2, 3), leading to the non-Darcian flow phenomenon.

Non-Darcian flow is a fundamental flow and transport problem, which has important implications to many natural and engineered processes in low-permeability porous media, such as geologic disposal of high-level nuclear waste, hydrocarbon energy extraction from shale formations, flow and contaminant transport in clay-rich aquifers, etc. (3, 11-17). Particularly, in a geologic nuclear waste repository, compacted bentonite clay has been proposed as a buffer material to fill the space between the host rock and a waste container (18, 19). Advanced understanding of non-Darcian flow in a bentonite buffer is critical to an accurate prediction of water migration and the related radionuclide transport in the near field of a nuclear waste repository over a time scale of 100,000 years. Specifically, the wetting process, caused by water imbibition from the surrounding rocks to the initially dry bentonite buffer, leads to bentonite saturation and swelling, thereby increasing in-situ stress and consequently improving the sealing effect of the buffer material.

Heat generated from nuclear waste will further affect water distribution and transfer in the bentonite buffer. A recent thermal-hydrological-mechanical (THM) simulation (20) showed that the temperature evolution in a bentonite-backfilled engineered barrier would range from less than 30 °C to ~ 90 °C over a time scale of 100,000 years. The temperature can be even higher depending on the configuration of waste emplacement. Non-Darcian flow is an important phenomenon to be considered in an overall THM modeling of a repository (20, 21). Numerous laboratory studies focused on non-Darcian flow and its dependence on hydrological and mechanical properties of the porous medium (15, 22, 23). However, our understanding of the role of temperature in such flow is very limited. Miller and Low (2) measured the threshold pressure gradient in a clay sample under 10.15 °C and 20.00 °C and found that the threshold pressure gradient decreased with increasing temperature. Zeng et al. (24) measured the threshold pressure gradient for oil flow in low-permeability sandstones under 70 °C and 90 °C and found a similar conclusion. Although these pioneering experimental studies provided valuable insights into the role of temperature, additional experimental data under multiple different temperatures and permeabilities are needed to unravel the underlying mechanism of the phenomenon. In addition, a theoretical framework for the interpretation of experimental data is generally missing.

In this work, we will first present a two-parameter model to describe the nonlinear relationship between flow velocity and pressure gradient for a non-Darcian flow in saturated low-permeability porous media based on the consideration of interfacial liquid-solid interactions (25). We then report our experimental results on the measurements of the threshold pressure gradient as a function of permeability and temperature. The data will be fitted to the two-parameter model. Finally, the model predictions of water flux under a wide range of permeabilities, pressure gradients, and temperatures will be presented and discussed.

## 2. Materials and Methods

### 2.1. Two-parameter model

**Figure 1A** illustrates the nonlinear relationship between the flow velocity and the hydraulic pressure gradient for a non-Darcian flow in a low-permeability porous medium. In this study, the threshold pressure gradient is defined as $J_t$, below which no fluid flow occurs (2, 24). $J_c$ is referred to as the critical gradient, which is the intersection of the x axis and the extension of the linear part of the curve (6). The definition of threshold pressure gradient ($J_t$) in our model is the same as other experimental studies such as Miller and Low (2), Zeng et al. (24), and Sanchez et al. (4). This is because $J_t$ is straightforward to measure in laboratory experiments if the accuracy of the equipment is sufficiently high.

We developed a continuum-scale, two-parameter model to account for non-Darcian flow in water-saturated low-permeability porous media based on the improvement of our previous study (25). This model hypothesizes that the porous medium is a bundle of circular tubes having the same diameter and fluid flow occurs inside the tubes. There is an immobile fluid layer at the liquid-solid interface on the inner walls of the circular tubes, which causes the threshold pressure gradient. This immobile layer thickness accounts for all combined effects that hinder fluid flow in saturated low-permeability porous media, such as strong liquid-solid interactions at solid surfaces and heterogeneity in pore size distribution. The thickness of the immobile layer, $h$, decreases exponentially as a function of the velocity gradient at the inner wall surface:

$$h = h_0 \exp\left(-a \frac{\partial u}{\partial r}\bigg|_{r=R}\right) \quad (1)$$

where $r$ is the distance from the tube center (m), $R$ is the tube radius (m), $h_0$ is the static immobile layer thickness (m) when the flow velocity is zero, and $a$ is characteristic time (s). $h_0$ and $a$ are the two parameters in the model, and their values are determined by fitting the model to experimental data, which will be discussed later. Equation 1 is a phenomenological correlation that aims to interpret the nonlinear relation between pressure gradient and flow velocity in saturated low-permeability porous media. For incompressible Newtonian laminar flow in a circular tube (26), the velocity gradient at the inner wall surface can be calculated as:

$$\frac{\partial u}{\partial r}\bigg|_{r=R} = \frac{R}{2\mu}\left|\frac{\partial p}{\partial x}\right| \quad (2)$$

where $\mu$ is fluid viscosity (Pa·s) and $\partial p / \partial x$ is pressure gradient (Pa/m). By substituting Equation 2 into Equation 1, one obtains:

$$h = h_0 \exp\left(-\frac{aR}{2\mu}\left|\frac{\partial p}{\partial x}\right|\right) \quad (3)$$

When the tube radius, $R$, is smaller than the static immobile layer thickness, $h_0$, no fluid flow occurs because the entire tube is blocked by the immobile layer. A minimum (i.e., threshold) value of the absolute value of pressure gradient, $|\partial p/\partial x|$, is required to reduce $h$ to a level smaller than $R$ in order to trigger flow in the tube. This threshold value is the threshold pressure gradient, $J_t$ (Pa/m). By substituting $h=R$ into the left-hand side of Equation 3, one obtains:

$$J_t = \begin{cases} \dfrac{2\mu}{aR}\ln\left(\dfrac{h_0}{R}\right) & h_0 \geq R \\ 0 & h_0 < R \end{cases} \tag{4}$$

which can be written in a compact form:

$$J_t = \max\left(0, \dfrac{2\mu}{aR}\ln\left(\dfrac{h_0}{R}\right)\right) \tag{5}$$

Equations 4 shows that $J_t=0$ when $R > h_0$, which indicates that no threshold pressure is needed to trigger the flow. This is consistent with our experimental observation that the threshold pressure gradient vanishes or decreases to an extremely low level which is non-detectible by laboratory equipment when the absolute permeability, which is positively correlated with the pore radius, $R$, is higher than a certain level.

The apparent permeability of the tube is calculated as:

$$k_p = \dfrac{1}{8}(R-h)^2, \qquad (R \geq h) \tag{6}$$

It should be noted that Equation 6 is valid only when the pressure gradient is higher than the threshold pressure gradient, which leads to $R \geq h$, thereby enabling flow to occur in the tube. Substituting Equation 3 into Equation 6, one obtains the formula for calculating the apparent permeability of the tube, which is a function of the pressure gradient:

$$k_p = \dfrac{1}{8}\left[R - h_0\exp\left(-\dfrac{aR}{2\mu}\left|\dfrac{\partial p}{\partial x}\right|\right)\right]^2, \qquad (|\partial p/\partial x| \geq J_t) \tag{7}$$

where $J_t$ is determined using Equation 4. Equation 7 can be substituted into the Darcy's law to calculate the Darcy velocity in the tube. It is clear that when $h_0 \geq R$, Equation 7 recovers a velocity-gradient relationship where $J_c > J_t > 0$, as shown in Figure 1A; when $h_0 < R$, $J_t = 0$ based on Equation 4 and thus Equation 7 reduces to the correlation of Swartzendruber (27), where $J_t = 0$ and $J_c > 0$; when $h_0 = 0$, Equation 7 reduces to the classic Darcy's law, where $J_t = J_c = 0$. This indicates that, despite the simple formulation, the two-parameter model accounts for a wide variety of Darcy and non-Darcian flows in porous media. In addition, when the magnitude of pressure gradient, $|\partial p / \partial x|$, is sufficiently large, the exponential term in Equation 7 becomes zero; in this scenario, the non-Darcian flow behavior is eliminated and thus the apparent permeability reduces to the absolute permeability, $k$ (i.e., $R^2/8$).

Because the porous medium is modeled as a bundle of these equally-sized circular tubes, the apparent permeability of the porous medium, $k_a$, is calculated as:

$$k_a = \frac{\phi}{8}\left[R - h_0 \exp\left(-\frac{aR}{2\mu}\left|\frac{\partial p}{\partial x}\right|\right)\right]^2, \qquad (|\partial p / \partial x| \geq J_t) \qquad (8)$$

where $\phi$ is porosity, defined as the ratio of total tube cross section area (i.e., total pore area) to total sample cross section area. The Darcy velocity through the porous medium is calculated using the Darcy's law:

$$U = \frac{-k_a}{\mu}\frac{\partial p}{\partial x} \qquad (9)$$

## 2.2. Laboratory experiments

Figure 1B illustrates the experimental setup. The threshold pressure gradient was measured as a function of absolute permeability under 20°C, 55°C, and 80°C. The sediment sample contained 10 wt.% bentonite clay and 90 wt.% fine sand. We mixed sand in the sample in order to accelerate the bentonite swelling process and enable the sediment structure to reach a steady state faster. The bentonite used in the experiments was the Wyoming bentonite clay, which had an average grain size smaller than 75 µm. The sand was rounded silica sand with an average grain size of 144.8 µm. The sediments were placed in a cylindrical core holder layer by layer to mitigate segregation of coarse and fine particles, and the core holder had a diameter of 25 mm and length of 60 mm. The core holder modified the confining pressure imposed on the sediments through a hydraulic pump to obtain different absolute permeabilities. We placed the core holder vertically in an oven and injected fluids from the bottom to eliminate preferential flows. The oven provided accurate temperature controls, and the injection fluid reservoir was also in the oven so that the fluid was in the equilibrium temperature. The injection fluid was a NaCl solution having ionic strength of 0.1 M, which is typical of that in natural host rocks (28). Because the core holder was vertically placed, the equivalent pressure gradient caused by gravity was incorporated into the total pressure gradient.

During the experiments, we first removed air from the sediment sample using a vacuum pump. We then fully saturated the sample by injecting the NaCl solution for multiple pore volumes and waited for an adequately long period of time to let clay swelling reach the steady state before the measurements. The solution was injected through the sample using a high-precision syringe pump at different flow rates. Particularly, we started from a high flow rate to eliminate the non-Darcian flow behavior. A differential pressure regulator measured the pressure difference between the two ends of the sample when the flow was in the steady state. We then reduced the injection flow rate sequentially and repeated the pressure difference measurement at each flow rate. The flow rate and pressure difference data measured within the Darcy regime were used to calculate the absolute permeability of the sample. When the flow rate decreased to an adequately low level, the flow entered the non-Darcian regime where the relationship between flow velocity and pressure gradient became nonlinear. The threshold pressure gradient, $J_t$, in the experiments was defined as the pressure difference normalized by the sample length when the flow rate reached 1 µL/min, which was the minimum flow rate that the syringe pump can accurately generate. When these injections were completed, we started a second round of injections from low flow rates to high flow rates. A second value of $J_t$ was read when the flow rate increased to 1 µL/min; the average of the two values was used as the final $J_t$ value. We conducted the second round of injections to investigate the path dependency of $J_t$ measurements. In practice, we found that the two measured $J_t$ values were similar, which implies that either increasing flow rates or decreasing flow rates will give the same $J_t$ value as long as the sediments are fully saturated.

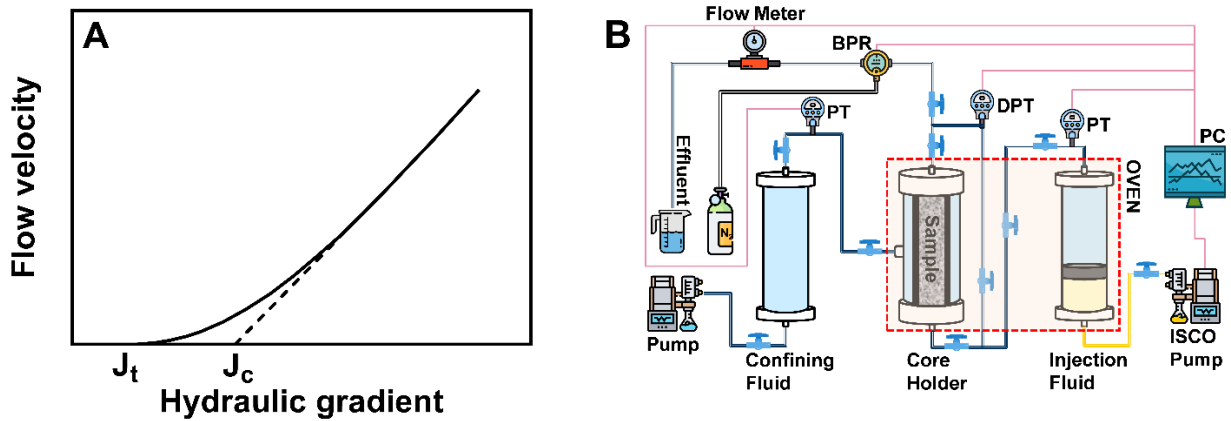

**Figure 1. A:** Nonlinear relationship between flow velocity and pressure gradient for non-Darcian flow in saturated low-permeability porous media. **B:** Schematic diagram of the experimental setup. BPR: back pressure regulator. DPT: differential pressure transducer. PT: pressure transducer.

## 2.3. Nonlinear least square fitting

In this study, the Gauss-Newton method, an iteration-based algorithm for nonlinear least square fitting, was used to fit laboratory measurement data to the two-parameter model to determine $h_0$ and $a$. Particularly, the vector of unknown parameters, $\beta$, which has two elements in this case (i.e., $h_0$ and $a$), is calculated by:

$$\beta^{n+1} = \beta^n + \left(\mathbf{J}^T\mathbf{J}\right)^{-1}\mathbf{J}^T\mathbf{r} \tag{10}$$

where $n$ indicates the iteration time step; $\mathbf{J}$ is the Jacobian matrix, in which each entry, $J_{ij}$, is the derivative of the model-predicted $J_t$ value (i.e., based on Equation 4) with respect to the j$^{th}$ unknown parameter at the i$^{th}$ measurement; $\mathbf{r}$ is the residual vector, in which each element, $r_i$, is the difference between the laboratory-measured and model-predicted $J_t$ values at the i$^{th}$ measurement.

## 3. Results and Discussion

### 3.1. Model fitting of experimental data

**Figure 2** illustrates laboratory-measured threshold pressure gradient as a function of absolute permeability under 20°C, 55°C, and 80°C. The absolute permeability was measured within the Darcy regime where the pressure gradient was sufficiently high, the immobile layer thickness becomes zero, and thus the velocity depends on the pressure gradient linearly. This is the first experimental study that measured the $J_t$-$k$ relationship in non-Darcian flow under several different temperatures. The threshold pressure gradient decreases with increasing permeability, which is consistent with other experimental studies (2, 5, 6, 29-32). Under the same permeability, a higher temperature results in a lower threshold pressure gradient, because an increased temperature would weaken the interaction energy between water molecules and clay surfaces (2). In addition, the power-law empirical correlation proposed by Liu and Birkholzer (6), $J_t = A \cdot k^{-B}$, was used to fit the measurement data shown in Figure 2. The fitted values of $B$ are 1.56, 2.08, and 2.55 under 20°C, 55°C, and 80°C, respectively. This indicates that the absolute value of the slope of the $J_t$-$k$ correlation increases with increasing temperature. In other words, the decline of threshold pressure gradient with increasing permeability is faster under a higher temperature. In order to explain these observations, the variations of $h_0$ and $a$ under different temperatures must be investigated to obtain mechanistic insights into the role of temperature on non-Darcian flow.

The measurement data were fitted to the two-parameter model (i.e., Equation 4) using nonlinear least square fitting. Because $k$ is the absolute permeability, $R$ and $k$ are related by $k=\phi R^2/8$ according to Equation 8. In addition, $k$ and $\phi$ are related based on the Kozeny-Carman correlation (33), $k = m\phi^\alpha$, where $m$ is a coefficient and $\alpha$ is the power-law exponent. Therefore, $\phi$ can written be as:

$$\phi = (k/m)^{1/\alpha} \qquad (11)$$

Substituting Equation 11 into $k=\phi R^2/8$, one can calculate $R$ as:

$$R = \sqrt{8 \cdot m^{1/\alpha} \cdot k^{(\alpha-1)/\alpha}} \qquad (12)$$

Based on the data fitting, we found m = $4.0\times10^{-15}$ and $\alpha = 2.5$, suggesting that $\phi$ and $k$ are positively correlated, which is consistent with previous studies (33, 34). Equations 11 and 12 indicate that the values of $\phi$ and $R$ can be determined given a particular value of $k$. In addition, fluid viscosity, $\mu$, is known at a given temperature, and in Equation 4 we use the $\mu$ value under the corresponding temperature, which suggests that the effect of temperature on fluid viscosity is accounted for in the model. Therefore, the only undetermined values in Equation 4 are the two parameters, $h_0$ and $a$. The goal of nonlinear least square fitting was to find the optimal values of $h_0$ and $a$ which gave the best fitting of the model to the laboratory measurement data.

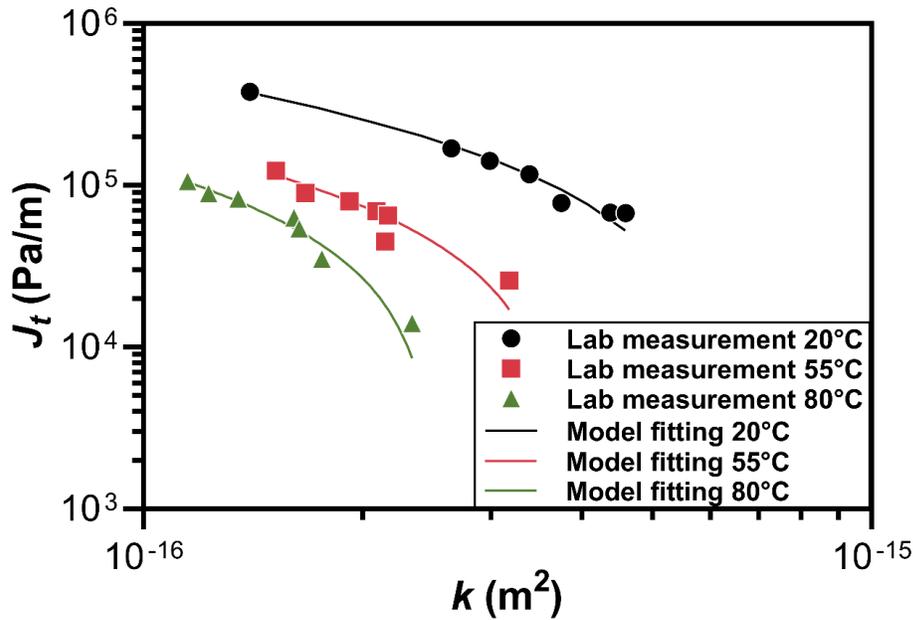

**Figure 2.** Threshold pressure gradient as a function of absolute permeability under 20°C, 55°C, and 80°C. The scatter data points are laboratory measurements, whereas the solid curves are obtained by fitting the two-parameter model to the data using nonlinear least square fitting.

The $R^2$ values of the nonlinear least square fitting were 0.991, 0.894, and 0.968 under 20°C, 55°C, and 80°C, respectively, which suggests that the two-parameter model successfully fits the laboratory data, as shown in Figure 2. Based on the fitting, the values of $h_0$ were found as 102.2 nm, 88.0 nm, and 78.2 nm under 20°C, 55°C, and 80°C, respectively; the values of $a$ were found as 0.037 s, 0.036 s, and 0.023 s under 20°C, 55°C, and 80°C, respectively. **Figure 3** illustrates $h_0$ and $a$ as a function of temperature. It is clear that both $h_0$ and $a$ decreased with an increasing temperature. Particularly, the decrease of $h_0$ under an increasing temperature is consistent with the hypothesis of our previous study (25), which suggests that an increasing temperature reduces the interaction energy near solid surfaces that hinders fluid movement in low-permeability porous media. In other words, an increasing temperature reduces the static thickness of the immobile layer. In addition, the lower value of $h_0$ under a higher temperature leads to the faster decline of threshold pressure gradient with increasing permeability, as demonstrated in Figure 2. This implies that under high temperatures the threshold pressure gradient vanishes or decreases rapidly to an extremely low level that is non-detectible by laboratory equipment when the permeability exceeds a certain level, which is consistent with our laboratory observations.

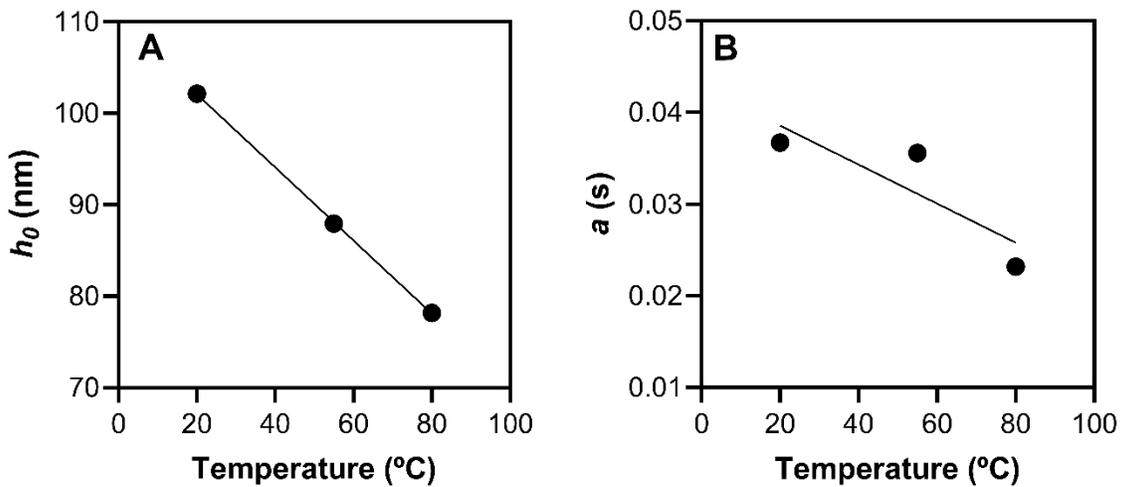

**Figure 3. A**: Static immobile layer thickness, $h_0$, and **B**: characteristic time, $a$, as a function of temperature. The solid lines are based on linear regression fitting.

## 3.2. Model extrapolation and prediction

The absolute permeability of an engineered clay barrier in a geological nuclear waste repository typically falls in the range between $10^{-20}$ m$^2$ and $10^{-18}$ m$^2$ (11, 35). We used the fitted values of $h_0$ and $a$ under 20°C, which were obtained by fitting laboratory data in the absolute permeability range between $10^{-16}$ m$^2$ and $10^{-15}$ m$^2$, and then extrapolated the two-parameter model to an absolute permeability level as low as $10^{-20}$ m$^2$. The model-predicted threshold pressure gradient at that permeability level was $5\times10^7$ Pa/m, which was in good agreement with other experimental data (2, 5, 29-32, 36). This implies that in the same clay material the values of $h_0$ and $a$ depend solely on temperature; when we increase the confining pressure to consolidate the clay and decrease the permeability, the same values of $h_0$ and $a$ can be used in the model to predict the threshold pressure gradient for the new permeability if the temperature stays the same. Note that such a high threshold pressure gradient (i.e., $5\times10^7$ Pa/m) is unlikely to achieve in a repository environment, implying that an advective water flow through an engineered barrier is not possible.

In addition to extrapolation in the dimension of absolute permeability, the two-parameter model can also be extrapolated in the dimension of pressure gradient. We substituted the values of $h_0$ and $a$ fitted under 20 °C into Equation 8 to determine the apparent permeabilities under a wide range of pressure gradient magnitude, $|\partial p / \partial x|$, which was then used in the Darcy's law (i.e., Equation 9) to predict the flow velocities under those pressure gradients. **Figure 4A** illustrates the model-predicted velocity-gradient curves under 20°C for three different absolute permeabilities. Similarly, we used the values of $h_0$ and $a$ fitted under 20°C, 55 °C, and 80 °C to predict the velocity-gradient curves under 20°C, 55 °C, and 80 °C, respectively, for similar absolute permeabilities, as shown in Figure 4B. The absolute permeability, $k$, was measured in the Darcy regime (i.e., linear section of the velocity-gradient curve), and different $k$ values were achieved by imposing different confining pressures on the clay sample. Therefore, the absolute permeability was an output of the experimental system, rather than an input. As a consequence, it was difficult to achieve the exactly same absolute permeability among different sets of experiments due to the different responses of sample's internal pore structure to the confining pressure under different temperatures. In practice, we consider the three absolute permeabilities in Figure 4B statistically equivalent. Figure 4 demonstrates that Equations 8 and 9 successfully predicted flow velocities in the Darcy regime by comparisons with laboratory measurement data. **Table 1** illustrates the model-predicted and laboratory-measured threshold pressure gradients for the five velocity-pressure gradient curves demonstrated in Figure 4. The agreement between model predictions and laboratory measurements indicates that nonlinear least square fitting successfully minimized the difference between the model and laboratory data, as shown in Figure 2.

Figure 4 illustrates the two-parameter model's extrapolation and prediction capability in the dimension of pressure gradient, which successfully describes the velocity-gradient curve in the nonlinear (i.e., non-Darcian) and linear (i.e., Darcy) regimes. Particularly, the threshold pressure gradient in the nonlinear regime determines the starting point of the curve, where flow starts to occur, whereas the linear section determines the absolute permeability. The extrapolation performance of the two-parameter model is attributed to the data fitting process which aims to determine the values of the two parameters (i.e., $h_0$ and $a$). Figure 2 shows that we used laboratory-measured $J_t$ and $k$ values to fit the values of $h_0$ and $a$. Particularly, $J_t$ determines the non-Darcian flow characteristics, whereas $k$ determines the Darcy flow characteristics because it is the absolute permeability measured in the linear section of the velocity-gradient curve. Therefore, the fitted values of $h_0$ and $a$ account for the flow properties in both the non-Darcian and Darcy regimes.

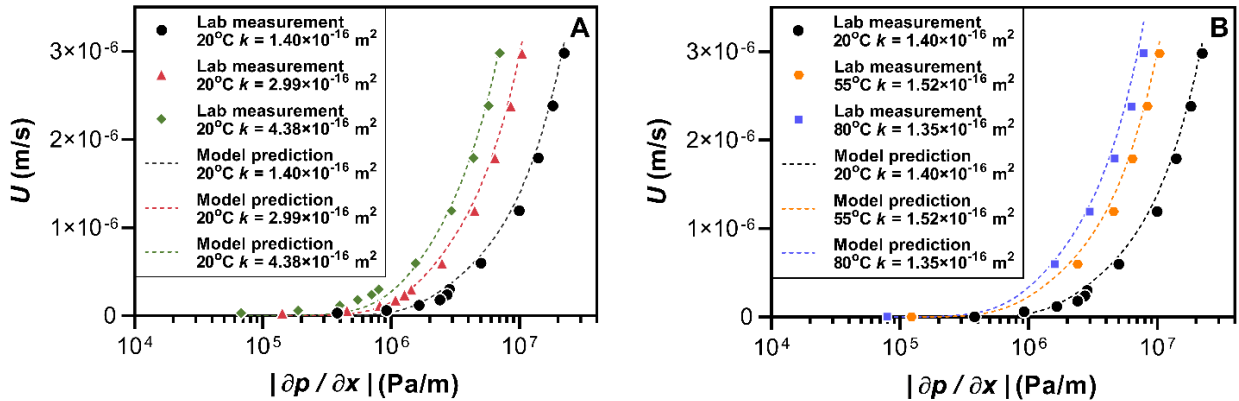

**Figure 4. A**: Flow velocity as a function of pressure gradient magnitude under 20 °C for three different absolute permeabilities. **B**: Flow velocity as a function of pressure gradient magnitude under 20 °C, 55 °C, and 80 °C for similar absolute permeabilities. The dash lines are model predictions whereas the scatter data points are laboratory measurements.

**Table 1.** Model-predicted and laboratory-measured threshold pressure gradients for the experiments demonstrated in Figure 4.

| Temperature (°C) | 20 | 20 | 20 | 55 | 80 |
|---|---|---|---|---|---|
| Absolute permeability (m$^2$) | 4.38×10$^{-16}$ | 2.99×10$^{-16}$ | 1.40×10$^{-16}$ | 1.52×10$^{-16}$ | 1.35×10$^{-16}$ |
| Model-predicted $J_t$ (Pa/m) | 6.17×10$^4$ | 1.46×10$^5$ | 3.74×10$^5$ | 1.15×10$^5$ | 8.00×10$^4$ |
| Lab-measured $J_t$ (Pa/m) | 6.77×10$^4$ | 1.42×10$^5$ | 3.80×10$^5$ | 1.23×10$^5$ | 8.30×10$^4$ |

In addition, Figure 4B demonstrates the two-parameter model's prediction capability in the dimension of temperature. It is clear that under a similar absolute permeability a higher temperature leads to a lower threshold pressure gradient. This indicates that under a higher temperature the flow enters the Darcy regime earlier with an increasing pressure gradient magnitude. In practice, the values of $h_0$ and $a$ under different temperatures need to be determined using data fitting to predict the velocity-gradient curve under each particular temperature.

## 4. Conclusions

In this work, we presented a novel two-parameter theoretical model that describes the nonlinear relationship between the flow velocity and the pressure gradient for a non-Darcian flow. We conducted well-controlled laboratory experiments to measure the threshold pressure gradient as a function of permeability and temperature. This is the first comprehensive experimental study that measured the $J_t$-$k$ curve under multiple different temperatures. The laboratory measurement data were then fitted to the two-parameter model, and the determined parameters (i.e., $h_0$ and $a$) provided mechanistic insights into the role of temperature on non-Darcian flow in saturated low-permeability porous media.

Particularly, the laboratory measurements showed that the threshold pressure gradient decreased with increasing permeability. Under the same permeability, a higher temperature resulted in a lower threshold pressure gradient. The decline of threshold pressure gradient with increasing permeability was faster under a higher temperature, which can be explained by the decreased static thickness of immobile layer (i.e., $h_0$) under a higher temperature. In addition, the two-parameter model has excellent extrapolation and prediction capabilities in the dimensions of absolute permeability, pressure gradient, and temperature. Specifically, the fitted values of $h_0$ and $a$ in the model account for flow properties in both the Darcy and non-Darcian regimes.

The novel two-parameter model and experimental data are valuable in fundamental studies of non-Darcian flow, which has critical implications to many engineered and natural processes associated with flow in low-permeability porous media, such as geological disposal of nuclear waste (6, 11, 15, 17), shale oil and gas recovery (37), and contaminant remediation in clay-rich formations (38).

## Acknowledgements

The authors are thankful to the financial support provided by the U.S. Department of Energy (DOE)'s Nuclear Energy University Program (NEUP) through the Award Number of DE-NE0008806. Data are available through: https://doi.org/10.5281/zenodo.6338520.